\documentstyle[12pt]{article}

\newcommand{\mysection}{\setcounter{equation}{0}\section}

\begin{document}
\vskip 0.2cm
\hfill{YITP-SB-00-68}
\vskip 0.1cm
\hfill{ANL-HEP-PR-00-113}
\vskip 0.1cm
\hfill{Version 2}
\vskip 1.0cm
\centerline{\large\bf {Variable flavor number schemes versus fixed order}}
\centerline{\large\bf {perturbation theory for charm quark electroproduction}}
\vskip 1.2cm
\centerline {A. Chuvakin, J. Smith }
\centerline{\it C. N. Yang Institute for Theoretical Physics}
\centerline{\it State University of New York at Stony Brook}
\centerline{\it Stony Brook, New York 11794-3840}
\vskip 0.5cm
\centerline {B. W. Harris }
\centerline{\it High Energy Physics Division}
\centerline{\it Argonne National Laboratory}
\centerline{\it Argonne, Illinois 60439}
\vskip 1.2cm
\centerline{November 2000}
\vskip 0.5cm
\centerline{\bf Abstract}
\vskip 0.3cm
Data for $D^{*\pm}(2010)$ meson electroproduction in the range 
$10<Q^2<1350 \; {\rm GeV}^2$ has recently been presented 
by the ZEUS collaboration at HERA.  We use these results together 
with previously published data for $Q^2 > 1 \; {\rm GeV}^2$ 
to test whether one can distinguish between different theoretical 
schemes for charm quark electroproduction.
We find that up to the largest $Q^2$ measured, it is not possible to 
make such a differentiation.  Then we point out the regions where 
differences between the various schemes arise.
\vfill

\mysection{Introduction}
\newcommand{\be}{\begin{eqnarray}}
\newcommand{\ee}{\end{eqnarray}}

Electromagnetic interactions have long been used to study both 
hadronic structure and strong interaction dynamics.  Examples include 
deep inelastic lepton-nucleon scattering, hadroproduction of 
lepton pairs, the production of photons with large 
transverse momenta, and various photoproduction processes involving 
scattering of real or very low mass virtual photons from hadrons.  
In particular, heavy quark production in deep inelastic 
electron-proton scattering is calculable in QCD and 
provides information on the gluonic content of the proton which 
is complementary to that obtained in direct photon production or 
structure function scaling violation measurements.  In addition, the scale 
of the hard scattering may be large relative to the 
mass of the charm quark, thus allowing one to study whether and 
when to treat the charm quark as a massless parton.  
It is this second aspect we wish to examine further in this paper.

The photon-gluon fusion mechanism is the simplest description 
of charmed quark electroproduction so that  
their production is assumed extrinsic, and their mass $m_c$ 
is retained throughout.  We call this description fixed order
perturbation theory (FOPT). It depends on a three-flavor set of
parton densities for the $u$, $d$, and $s$ quarks together
with a corresponding gluon density. Calculations for rates and 
single particle inclusive distributions are available 
to next-to-leading order (NLO) in \cite{lrsn}.
These calculations were later redone to cover fully 
differential production \cite{hs},
and decays into hadronic or semileptonic final states \cite{hs1}.
This framework generally provides a very good description of the ZEUS 
\cite{ZEUS} and H1 data \cite{H1} on the differential distributions
for $D^{*\pm}(2010)$ electroproduction.  Updated analyses now exist 
from H1 \cite{H1_osaka} and ZEUS \cite{osaka}.
The ZEUS data \cite{osaka} now extend
up to $Q^2 \approx 1000 \; {\rm GeV}^2$.
Since the FOPT results in NLO are very stable under
scale changes it has been advocated that a three-flavor
description should be the best one to fit the
data \cite{grs,harris}. This is the reason that the GRV98 
leading order (LO) and NLO density sets 
\cite{grv98} only contain three flavors.

Other descriptions of charm quark electroproduction have been used. 
One, which describes the charm quark as a massless parton 
density $c(x,\mu^2)$, with the boundary condition $c(x,\mu^2) =0 $ 
for $\mu \le m_c$, is expected to be more appropriate at large 
$Q^2$. 
This scheme has generally been used by groups which 
fitted parton densities to data and is called the zero-mass variable 
flavor number scheme (ZM-VFNS). 
The transition from a three-flavor parton density
set to a four-flavor set can be made on purely theoretical 
grounds by evaluating appropriate massive and massless operator matrix 
elements containing heavy quark loops
in the operator product expansion and then absorbing the terms 
containing $\ln^i (Q^2/m_c^2)$ into 
the definition of four-flavor parton densities \cite{bmsn1}, \cite{chsm}.
The resummation of the above logarithms is incorporated into the 
boundary conditions on the $c-$quark density as well as the other
four-flavor quark and gluon densities. In particular if one does this
at the scale where $\mu^2= Q^2 = m_c^2$ then all the logarithmic
terms in the operator matrix elements vanish and only the non-logarithmic
terms are included in the boundary conditions. 
Since a charm quark density is a parton model concept 
the QCD perturbation series then starts with $\alpha_s^0$ 
coefficient functions. The lowest order photon-gluon fusion reaction then has
$O(\alpha_s)$ coefficient functions. The 
NLO corrections contain $O(\alpha_s^2)$ coefficient functions.
When the resulting four-flavor parton densities are convolved with 
the massless coefficient functions in \cite{zn}
one obtains predictions for the 
charm content in the deep inelastic structure functions. 
One expects this four-flavor ZM-VFNS description
to be better than the FOPT one at large scales since it resums the 
terms in $\ln^i(Q^2/m_c^2)$. 

Another approach, which is even more ambitious, is a scheme designed to 
interpolate between the FOPT result at low scales 
and the ZM-VFNS result at large scales. 
In these variable flavor number schemes (VFNS) one hopes
to provide a unified framework for all scales.  Unfortunately there is 
no unique prescription for a VFNS and several have been constructed. 
The differences between them are due to two inputs.
The first is the mass factorization 
procedure carried out before the large logarithms can be resummed, 
namely should one retain massive or massless charmed quarks in the 
coefficient functions, which are convolved with either
the three-flavor or four-flavor parton densities.
The second is the matching condition imposed on 
the charmed quark density, namely how does it vanish 
in the threshold region of the electroproduction 
process, where the partonic subenergy is approximately $4 m_c^2$ . 
Variable flavor number schemes are presently
available to $O(\alpha_s)$ in \cite{acot,cteq5,mrst98,kos} and 
to $O(\alpha_s^2)$ in \cite{bmsn2,csn}, called BMSN and CSN, respectively, 
in this paper.  
The latter schemes require the parton densities provided in \cite{chsm}.
Review articles and discussions about VFNS schemes are available 
in \cite{ns1,thro,col,astw}.


In Sec.\ II we give a short discussion of the BMSN and CSN descriptions
for charmed quark electroproduction and then compare theoretical
predictions with differential distribution data from H1 \cite{H1_osaka} 
and ZEUS \cite{ZEUS,osaka}.


\pagestyle{myheadings}  
\mysection{Comparison}

The reaction under consideration is heavy quark $Q$ 
production via neutral-current electron-proton scattering
\begin{eqnarray}
e^-(l) + P(p) \rightarrow e^-(l') + Q(p_1) + X.
\end{eqnarray}
We concentrate on the case where $Q$ is a charm quark
with mass $m_c = 1.4$ GeV.
When the momentum transfer squared $Q^2 = - q^2 >0 $
($q=l-l')$ is not too large, $Q^2 \ll M_Z^2$, the contribution
from virtual Z-boson exchange is small compared to that of 
virtual-photon exchange.  For example, using the leading order 
Monte Carlo program AROMA \cite{aroma}, at $Q^2=1000\; {\rm GeV}^2$ 
we find the Z-boson exchange contribution is a factor of 
$100$ smaller than the photon exchange contribution.

The charm quark cross section can be written 
in terms of the structure functions 
$F_2^c(x,Q^2,m_c^2)$ and $F_L^c(x,Q^2,m_c^2)$
as follows:
\begin{eqnarray}
\frac{d^2\sigma}{dydQ^2} = \frac{2\pi\alpha^2}{yQ^4}
\{[1+(1-y)^2]
F_2^c(x,Q^2,m_c^2) - y^2 F_L^c(x,Q^2,m_c^2)\}\,,
\end{eqnarray}
where $x=Q^2/2p \cdot q$ and $y=p \cdot q/p \cdot l$ are the
usual Bjorken scaling variables and $\alpha$ is the 
electromagnetic coupling. The scaling variables are 
related to the square of the center-of-momentum energy of the 
electron-proton system $S=(l+p)^2$ {\em via} $xyS=Q^2$. 
The total cross section is given by \cite{schuler}
\begin{eqnarray}
\sigma = \int^1_{4 m_c^2/S} \, dy
\int^{yS-4m_c^2}_{m_e^2y^2/(1-y)} \, dQ^2
\left( \frac{d^2\sigma}{dy dQ^2} \right)\,,
\end{eqnarray}
where $m_e$ is the electron mass. In deriving Eq.\ (2.2) one
integrates over the azimuthal angle between the plane 
containing the incoming and outgoing electrons and the plane
containing the incoming proton and the outgoing charm
quark.  

Experimentally it is the decay products of charmed hadrons 
that are observed.  The H1 and ZEUS groups measure $D^{*\pm}(2010)$ 
production.  We assume a Peterson {\em et al}.\ \cite{Peterson}
fragmentation function to model the nonperturbative transition from 
charmed quark to hadron.  
The cross section for $D^*$ production is then obtained by
convolving the charm quark cross section Eq.\ (2.3) with the 
fragmentation function
\begin{eqnarray}
D(z) = \frac{N}{z[1-1/z-\epsilon/(1-z)]^2}
\end{eqnarray}
where $N$ is fixed such that $D(z)$ is normalized to unity 
once the parameter $\epsilon=0.035$ \cite{no} is fixed.
The normalization of the cross section is then given
by the charm fragmentation probability which we take as  
$P(c\rightarrow D^*)=0.235$ \cite{prob}.

The H1 collaboration has recently \cite{H1_osaka} measured 
$D^{*\pm}$ production for $1 < Q^2 < 100 \; {\rm GeV}^2$
and $0.05 < y < 0.7$ and quote 
a cross section in the region $1.5 < p_T(D^*) < 15 $ GeV 
and $|\eta(D^*)| < 1.5$ of
\begin{eqnarray}
\sigma(e^+ p \rightarrow e^+ D^{*\pm} X) =
8.37 \pm 0.41 ({\rm stat.})^{+1.11}_{-0.82} 
({\rm syst.}) \; {\rm nb}\,.
\end{eqnarray}
The data came from the 1996-97 run with 
proton energy 820 GeV and positron energy 27.5 GeV
(18.6 pb$^{-1}$). 

The ZEUS collaboration has recently \cite{osaka} measured 
$D^{*\pm}$ production for $Q^2>10 \; {\rm GeV}^2$
and $0.04 < y < 0.95$ and quote 
a cross section in the region $1.5 < p_T(D^*) < 15 $ GeV 
and $|\eta(D^*)| < 1.5$ of
\begin{eqnarray}
\sigma(e^+ p \rightarrow e^+ D^{*\pm} X) =
2.33 \pm 0.12 ({\rm stat.})^{+0.14}_{-0.07} 
({\rm syst.}) \; {\rm nb}\,.
\end{eqnarray}
The data came partly from the 1999-2000 run with 
proton energy 920 GeV (45.0 pb$^{-1}$) 
and partly from the 1995-1997 run with 
proton energy 820 GeV (37.6 pb$^{-1}$). In both cases
the positron beam energy was 27.6 GeV. 
They demonstrated \cite{osaka} that the predictions from HVQDIS \cite{hs1}, 
which is based on FOPT, agree with their $D^{*\pm}$ electroproduction data 
up to the highest measured value of $Q^2 \approx 1350$ ${\rm GeV}^2$. 

Previously in \cite{ZEUS} they presented the 1996-1997 positron production 
data
(37 pb$^{-1}$) for $D^{*\pm}$ in the range
$1 < Q^2 < 600$ ${\rm GeV}^2$ and $0.02 < y < 0.7$ in the same kinematic
range 
$1.5 < p_T(D^*) < 15 $ GeV and $|\eta(D^*)| < 1.5$
with the cross section 
\begin{eqnarray}
\sigma(e^+ p \rightarrow e^+ D^{*\pm} X) =
8.31 \pm 0.31 ({\rm stat.})^{+0.30}_{-0.50} 
({\rm syst.}) \; {\rm nb}\,.
\end{eqnarray}
Therein \cite{ZEUS} they also concluded that
the HVQDIS \cite{hs1} results agree with their data, apart from a 
distortion of the pseudo-rapidity distribution.  
This was attributed to a beam drag effect \cite{beam}, 
which was estimated by Monte Carlo \cite{rap}.
Still FOPT seemed to be the best model to fit their data.  

The BMSN \cite{bmsn2} and CSN \cite{csn} variable flavor number 
schemes were constructed so that the charm quark structure functions 
$F_2^c(x,Q^2,m_c^2,\mu^2)$ and $F_L^c(x,Q^2,m_c^2,\mu^2)$
are numerically equal to the corresponding FOPT results
at the scale $\mu^2 = m_c^2 = Q^2= 1.96\; {\rm GeV}^2$, 
so the differences between the them could be monitored at 
higher scales.   For this reason we chose the scale 
$\mu^2 = m_c^2 + \frac{1}{2} Q^2 (1 - m_c^2/Q^2)^2$ and set the
charm density to zero when $\mu^2 < m_c^2$. Also we used the exact
solution of the differential equation for the QCD
running coupling $(\alpha_s)$ as well as an electromagnetic
running coupling $(\alpha)$.
The QCD expansion was truncated at $\alpha_s^2$. 
Therefore we were careful to construct the structure
functions in the FOPT, BMSN and CSN schemes
according to the symbolic formula
\begin{eqnarray}
F (x,Q^2,m_c^2 ) &=& f(LO) \otimes C(LO) 
\nonumber \\
&+& \alpha_s [ f(NLO) \otimes C(LO) + f(LO) \otimes C(NLO)]
\nonumber \\
&+& \alpha_s^2 [ f(LO) \otimes C(NNLO) + f(NLO) \otimes C(NLO)
\nonumber \\
&+& f(NNLO) \otimes C(LO)] \,,
\end{eqnarray}
where the $\otimes$ symbol refers to the convolution integral and the
parton densities $f$ and coefficient functions $C$ are taken in either
LO, NLO or next-to-next-to-leading order (NNLO) perturbation theory. 
Note that this result is different from the usual FOPT prescription
which is based on expressions like 
\begin{eqnarray}
F (x,Q^2,m_c^2 ) &=& [ f(LO) + \alpha_s f(NLO) + \alpha_s^2 f(NNLO)] 
\nonumber \\
& \otimes & [ C(LO) + \alpha_s C(NLO) + \alpha_s^2 C(NNLO)] \,.
\end{eqnarray}
These prescriptions retain terms which are even higher order 
in $\alpha_s$.
Normally it does not matter if such terms are
retained because they are numerically unimportant at large $Q^2$.
However such terms are numerically significant at small $Q^2$ 
and ruin the cancellations among the various terms in our formulae
for the structure functions in the three schemes with the 
result that they do not agree numerically at $Q^2 = \mu^2 = m_c^2$.
Therefore we have to use Eq.\ (2.8) and not Eq.\ (2.9).
Even our FOPT (extrinsic) expression, called EXACT in this paper,
only retains the second and third sets of terms in Eq.\ (2.8)
and agrees with the corresponding results from the
(appropriately modified) HVQDIS code.

The difference between the BMSN and CSN schemes is that the former 
has $m_c=0$ in the heavy quark coefficient functions while the latter 
retains terms containing $m_c$.
We refer the reader to \cite{csn} for
more details, in particular the definition of the collinear safe 
inclusive structure functions and the contributions which are 
incorporated into the light mass ($u,d,s$) contributions to
the coefficient functions.
Our previous theoretical results showed that
differences between the EXACT, BMSN and CSN schemes in LO perturbation theory
diminish substantially in NLO perturbation theory. Such differences are
more apparent for b-quark electroproduction in \cite{csn1}
but there are no events yet.  

The aim of this paper is to compare our results with the data. 
Since there are no VFNS schemes available 
for differential distributions
in the transverse momentum and rapidity of the $D^*$ meson in
$O(\alpha_s^2)$ we make the assumption that the 
experimental acceptances do not differ much between these schemes and FOPT.
We have therefore recalculated the experimental acceptances
from the HVQDIS program with the above scale choice, 
running coupling constant and GRV98 \cite{grv98} three-flavor parton density
set. This is appropriate for the FOPT result, which we called EXACT in
our papers on the BMSN and CSN schemes. 
Our acceptances in $Q^2$ are slightly modified 
from those used by the Collaborations. 
The acceptances in $Q^2$, from integrating Eq.\ (2.2)
over $0.04 < y < 0.95$, $0.05 < y < 0.7$, or $0.02 < y < 0.7$ are 
nearly identical, so we do not distinguish between them.
The corresponding acceptances in $x$, however, from integrating
Eq.\ (2.2) over $10 < Q^2 < 1350$ or $1 < Q^2 < 600$ and 
the corresponding $y$ ranges are different and we distinguish between them. 
We start with the recent data from the Osaka meeting \cite{H1_osaka,osaka}. 
The results for the ratio 
$\sigma({\rm cuts})/\sigma({\rm no}\,\,{\rm cuts})$ 
are presented in Figs.\ 1 and 2 plotted versus ${\rm log}_{10}Q^2$ 
and ${\rm log}_{10}x$ respectively. These plots demonstrate
the large corrections necessary to include the experimental
acceptances.  The corrections were applied to the corresponding 
differential cross sections calculated from the structure functions 
given in the CSN \cite{csn} and BMSN \cite{bmsn2} papers. 
Here we used our own set of densities \cite{chsm}
which are based on the three-flavor GRV98 densities at 
scales below $\mu = m_c$, but which incorporate the discontinuity 
across the $c-$flavor threshold at $\mu = m_c$ to define a four-flavor 
set both in $O(\alpha_s)$ and in $O(\alpha_s^2)$ together with their 
subsequent evolution to higher scales with NLO splitting functions.  

The resulting differential cross sections in ${\rm log}_{10}Q^2$
are compared with the H1 and ZEUS data in Figs.\ 3 and 4.  
We see from Fig.\ 3 that the FOPT is a good fit
to the data at large $Q^2$.  This is in agreement with the conclusions
of the ZEUS collaboration in \cite{osaka}.  It is difficult 
to distinguish the BMSN and CSN results from the FOPT ones because there 
is only a $4\%$ difference even at this large $Q^2$.
Clearly it will take a substantial increase in the number of events
to distinguish between the schemes at large $Q^2$. 
All we can say at present is that the terms containing powers
of $\ln(Q^2/m_c^2)$ do not seem to lead to different predictions.

One can see from the semi-logarithmic plot in Fig.\ 4 that all curves
meet at $Q^2 = m_c^2 = 1.96\; {\rm GeV}^2$, which is expected 
from the construction of the BMSN and CSN schemes.  There are
differences between the three schemes in the region of small $Q^2$, however 
the currently available data is unable to resolve them.
We understand that the events with $Q^2 < 10$ ${\rm GeV}^2$ and
with $Q^2 > 10$ ${\rm GeV}^2$ are measured in different
regions of the ZEUS detector and the events accumulated in 1999-2000
in the former region have not been analyzed.  Note also that the 
bin widths in this region are not the same.
More data for small $Q^2$ would clearly be very useful.

The resulting differential cross sections in ${\rm log}_{10} x$
are compared with the new ZEUS data \cite{osaka} in Figs.\ 5 and 6.
We see from Fig.\ 5 that there is good agreement over
a wide range in $x$. The semi-logarithmic plot in Fig.\ 6 shows
a small disagreement between the FOPT theory result and
the data in the region $x \approx 10^{-3}$.  However the normalization is 
determined mainly by the magnitude of differential cross section at
the lowest measured point in $Q^2$, which is precisely
where additional data is required.

Integrations over the theoretical results displayed in Figs.\ 3 - 6 for 
$10 < Q^2 < 1350$ ${\rm GeV}^2$ and $0.04 < y < 0.95$
yield 2.86 nb, 2.51 nb and 2.48 nb for the FOPT, BMSN
and CSN schemes respectively. 
The latter two results are within the error
bars of the experimental result in Eq.\ (2.6) while the FOPT
result is slightly higher.

The previous published ZEUS data in \cite{ZEUS} had different cuts
in $Q^2$ and $y$, namely $1 < Q^2 < 600$ ${\rm GeV}^2$
and $0.02 < y < 0.7$ which affect the normalization
of the corresponding $x$ distribution. Therefore we reran the acceptance in
${\rm log}_{10} x$ from the HVQDIS program and it is shown in Fig.\ 7.
We then applied the same acceptance to the other programs. The BMSN
and CSN results between $1 < Q^2 < 1.96$ ${\rm GeV}^2$ are set 
equal to the EXACT result. 
Our results are compared to the data in Figs.\ 8 and 9. 
The overall shape and normalization are well described. Integration over
the results in Fig.\ 8 yield 9.29 nb, 8.43 nb and 8.55 nb for 
the FOPT, BMSN and CSN schemes respectively compared to the experimental
results in Eq.\ (2.5) and (2.7).

We have run our computer codes in other ranges of the variables
${\rm log}_{10} Q^2$ and ${\rm log}_{10} x$ to find 
where differences between the three schemes might be measurable. 
As an illustration we show in Fig.\ 10 a contour plot of the ratio of the 
BMSN double differential cross section 
divided by the FOPT double differential cross section plotted versus
these variables. Contour lines are drawn where this ratio is 1,
1.5, 2, 2.5 and 3.  The ratio increases as $Q^2$ increases for
fixed $x$. Note that no acceptance corrections in $p_T$ or $\eta$
have been applied to the ratio in this figure. 
One sees that the region of large $Q^2$ and large $x$ 
must be probed to find significant differences between FOPT and the 
variable flavor number schemes.  Roughly speaking one needs $x > 0.2$ and 
$Q^2 > 100$ ${\rm GeV}^2$. 
In fact Fig.\ 5 in \cite{js},
which only shows the $Q^2$ dependence of the structure function
$F_2^c(x,Q^2,m_c^2)$ at fixed values of $x$,    
already illustrates the kind of differences one can expect in this region.

Finally we remark that as far as the FOPT result is concerned the
standard version of HVQDIS uses the scale $\mu^2 = Q^2 + 4m_c^2$.
This increases the scale in the running coupling.
Also it uses the second and third lines of Eq.\ (2.9) with 
all parton densities set to their three flavor NLO values. These
standard settings alter slightly the above acceptance curves.
The net effect of both changes is approximately a ten percent reduction of 
the FOPT results (usgin GRV98) for the differential distribution 
in ${\rm log}_{10} Q^2$ at the smallest $Q^2$, which is within the 
present experimental errors.

To summarize we have made a first comparison between the FOPT,
BMSN and CSN descriptions for $D^{*\pm}$ electroproduction.
We have observed that the three schemes give nearly identical 
predictions up to the highest $Q^2$ measured.  It is therefore 
difficult to distinguish between the various schemes on the 
basis of a data comparison.  The small scale dependence of the FOPT 
result indicates that there is no sign that the terms containing 
powers of $\ln(Q^2/m_c^2)$ destroy the convergence of the
QCD perturbation expansion and that one is 
forced to switch to a variable flavor number scheme like the BMSN or CSN. 
In fact they all provide a good description of the data for the
differential distributions in $Q^2$ and $x$.
At small $Q^2$ there is a chance to distinguish between the schemes 
(say for $2 < Q^2 < 20$ in ${\rm GeV}^2$).
The comparisons in the case of the $x$-distributions are not 
conclusive due to the correlations with the points in small $Q^2$. 
It will be interesting to see what happens when more events are 
collected so that the error bars are reduced.

\vspace*{4ex}
We would like to acknowledge discussions with W.\ L.\ van Neerven
on the results presented above, and thank Jos\'{e} Repond for 
discussions about the ZEUS data and comments on the text.
The work of A. Chuvakin and J. Smith was supported in part by
the National Science Foundation Contract PHY-9722101.  The work 
of B. Harris was supported by the U.S. Department of Energy, 
High Energy Physics Division, under contract W-31-109-Eng-38.


%

\centerline{\bf \large{Figure Captions}}
\begin{description}
\item[Fig. 1.]
The ratio $\sigma({\rm cuts})/\sigma({\rm no}\,\,{\rm cuts})$
for the acceptance in $Q^2$ plotted versus ${\rm log}_{10} Q^2$.
\item[Fig. 2.]
The ratio $\sigma({\rm cuts})/\sigma({\rm no}\,\,{\rm cuts})$
for the acceptance in $x$ plotted versus ${\rm log}_{10} x$.
\item[Fig. 3.]
The combined Osaka H1 and ZEUS and published 
ZEUS data for $d\sigma/d\,{\rm log}_{10}{Q^2}$ in nb 
for deep inelastic production of $D^{*\pm}$ mesons.
The dashed line is the NLO EXACT result from HVQDIS, 
(which coincides with the FOPT result), the
dotted line is the result from the BMSN scheme
and the dot-dashed line is the result from the CSN scheme.
\item[Fig. 4.]
Same as Fig.\ 3 displayed on a semi-logarithmic plot.
\item[Fig. 5.]
The Osaka ZEUS data for $d\sigma/d\,{\rm log}_{10}{x}$ in nb 
for deep inelastic production of $D^{*\pm}$ mesons.
The dashed line is the NLO EXACT result from HVQDIS, 
(which coincides with our FOPT result), the
dotted line is the result from the BMSN scheme
and the dot-dashed line is the result from the CSN scheme.
\item[Fig. 6.]
Same as Fig.\ 5 displayed on a semi-logarithmic plot.
\item[Fig. 7.]
The ratio $\sigma({\rm cuts})/\sigma({\rm no}\,\,{\rm cuts})$
for the acceptance in $x$ plotted versus ${\rm log}_{10} x$.
\item[Fig. 8.]
The published ZEUS data for $d\sigma/d\,{\rm log}_{10}{x}$ in nb 
for deep inelastic production of $D^{*\pm}$ mesons.
The notation follows Fig.\ 5.
\item[Fig. 9.]
Same as Fig.\ 8 displayed on a semi-logarithmic plot.
\item[Fig. 10.]
Ratio of the double differential cross sections
in ${\rm log}_{10} Q^2$ and ${\rm log}_{10} x$ for the 
BMSN scheme divided by the FOPT result. The contour lines
are for the ratio 1, 1.5, 2, 2.5 and 3 in the order
of increasing $Q^2$ for fixed $x$.
\end{description}

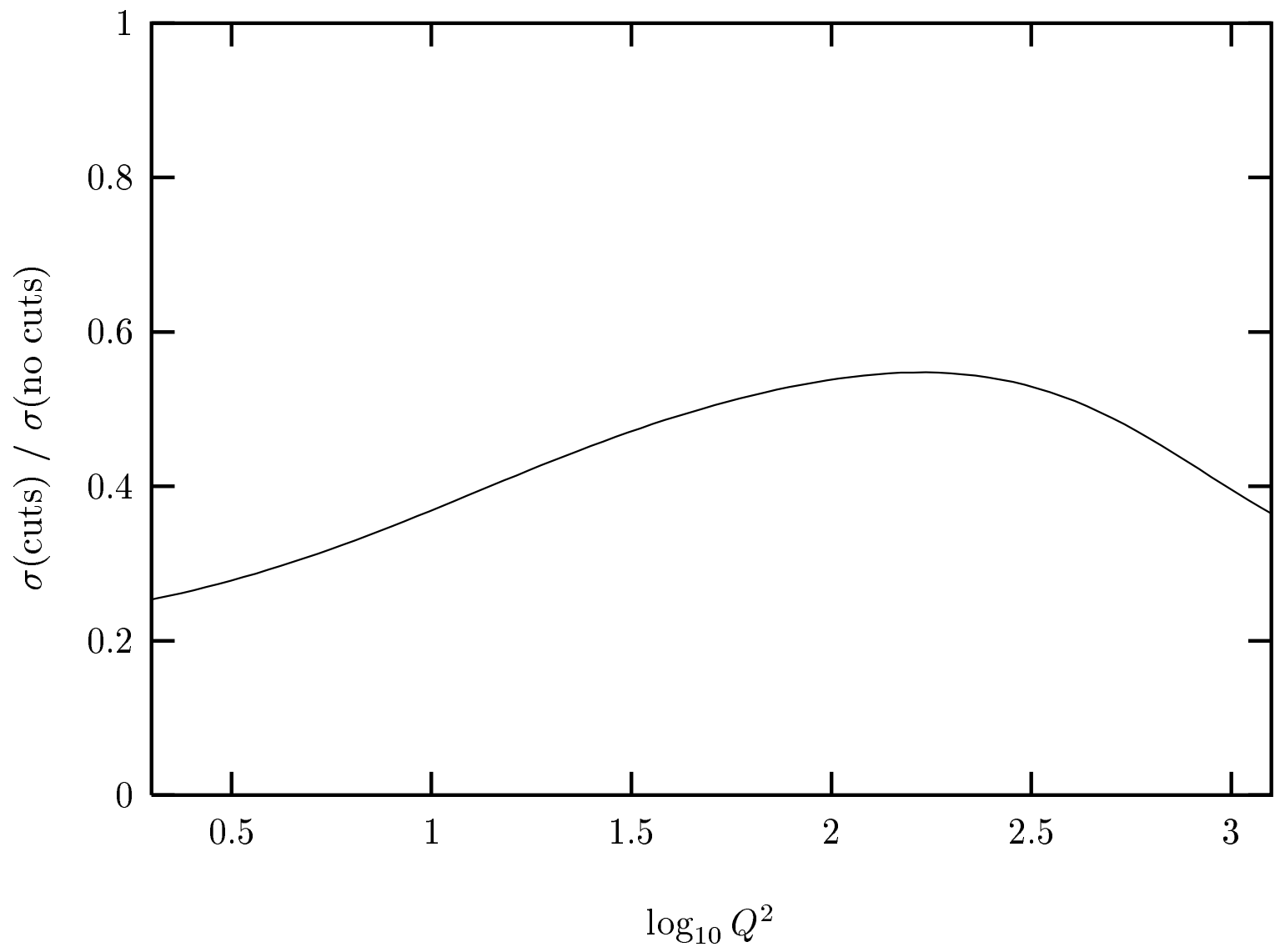
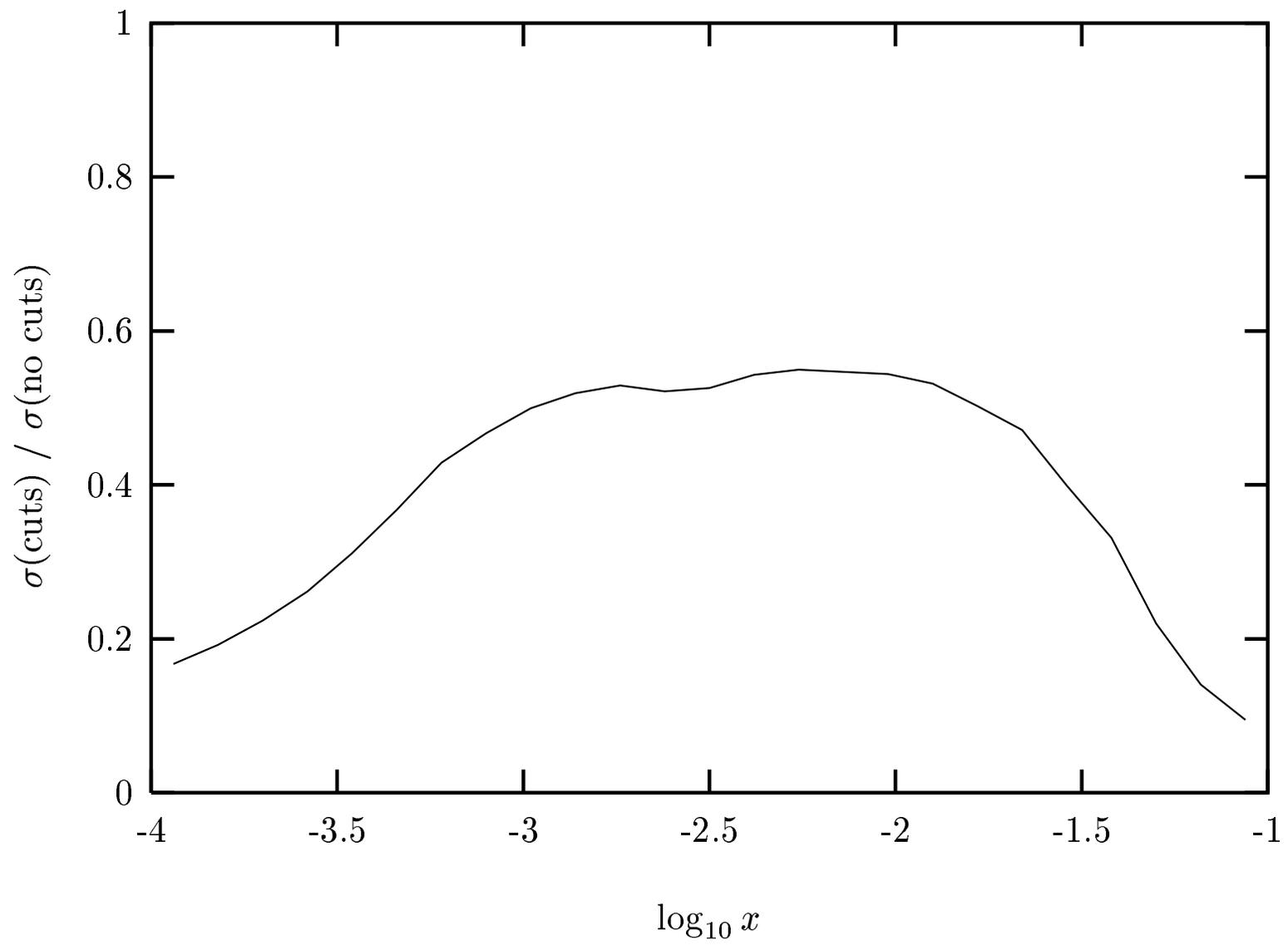
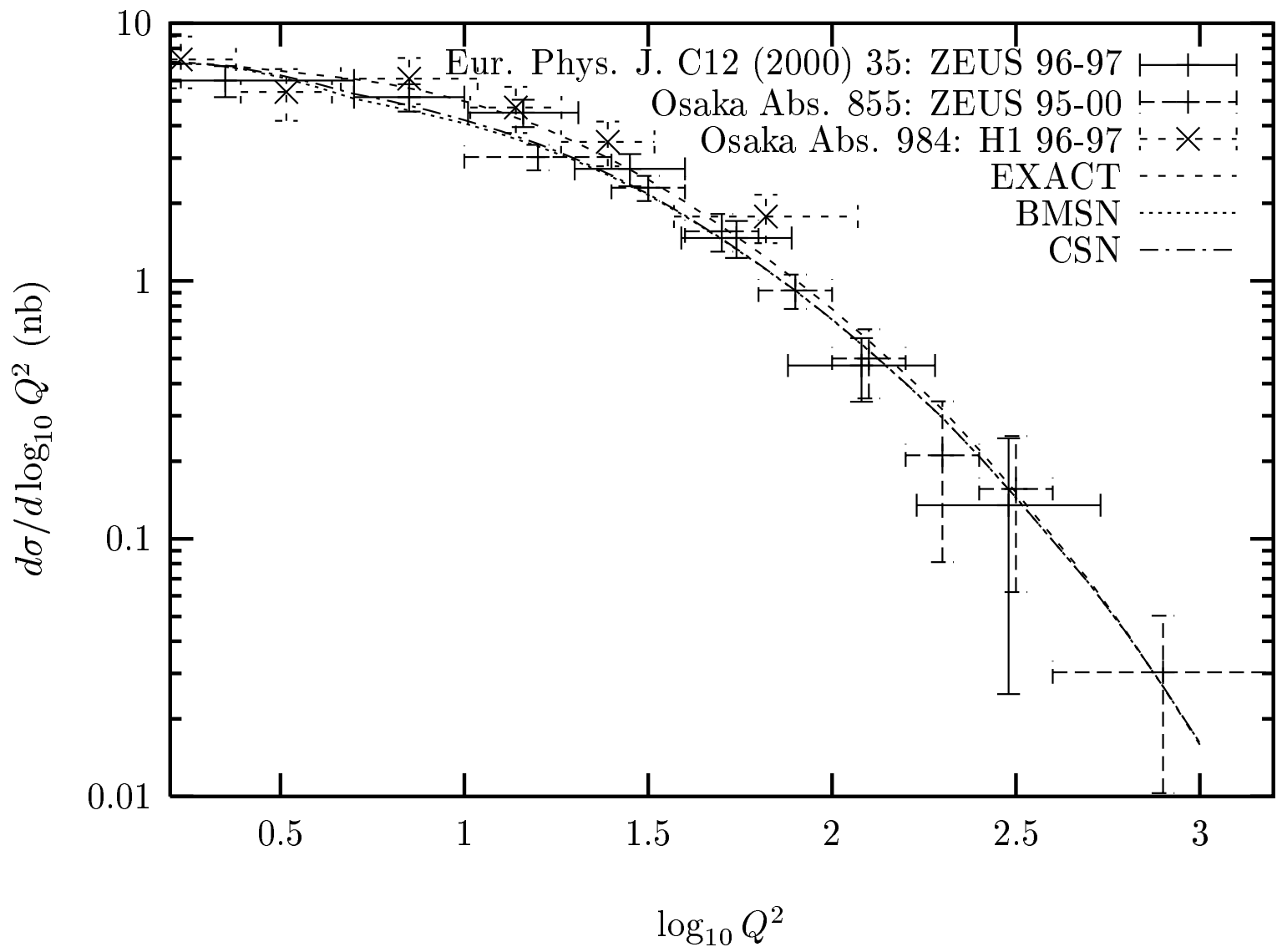
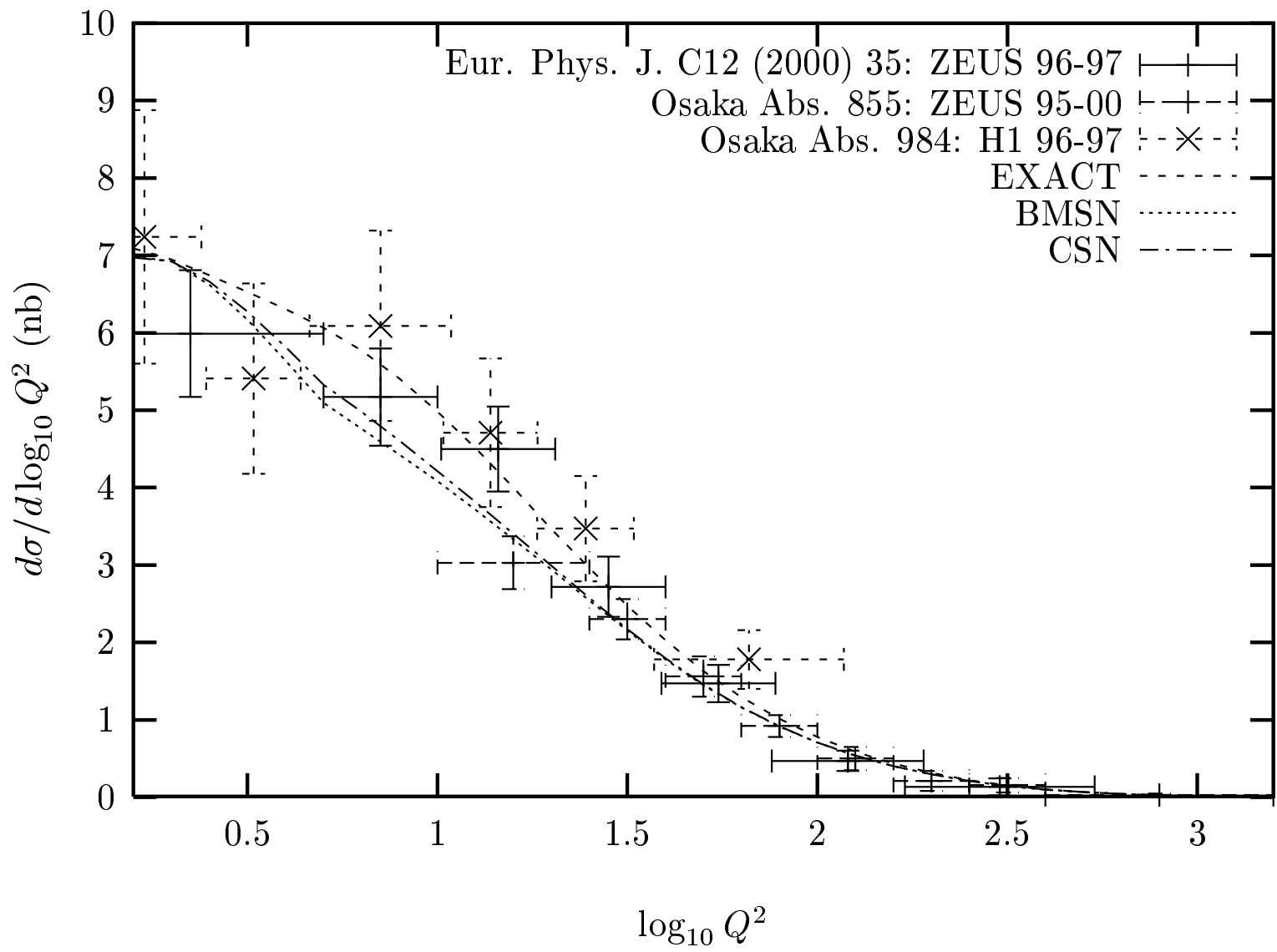
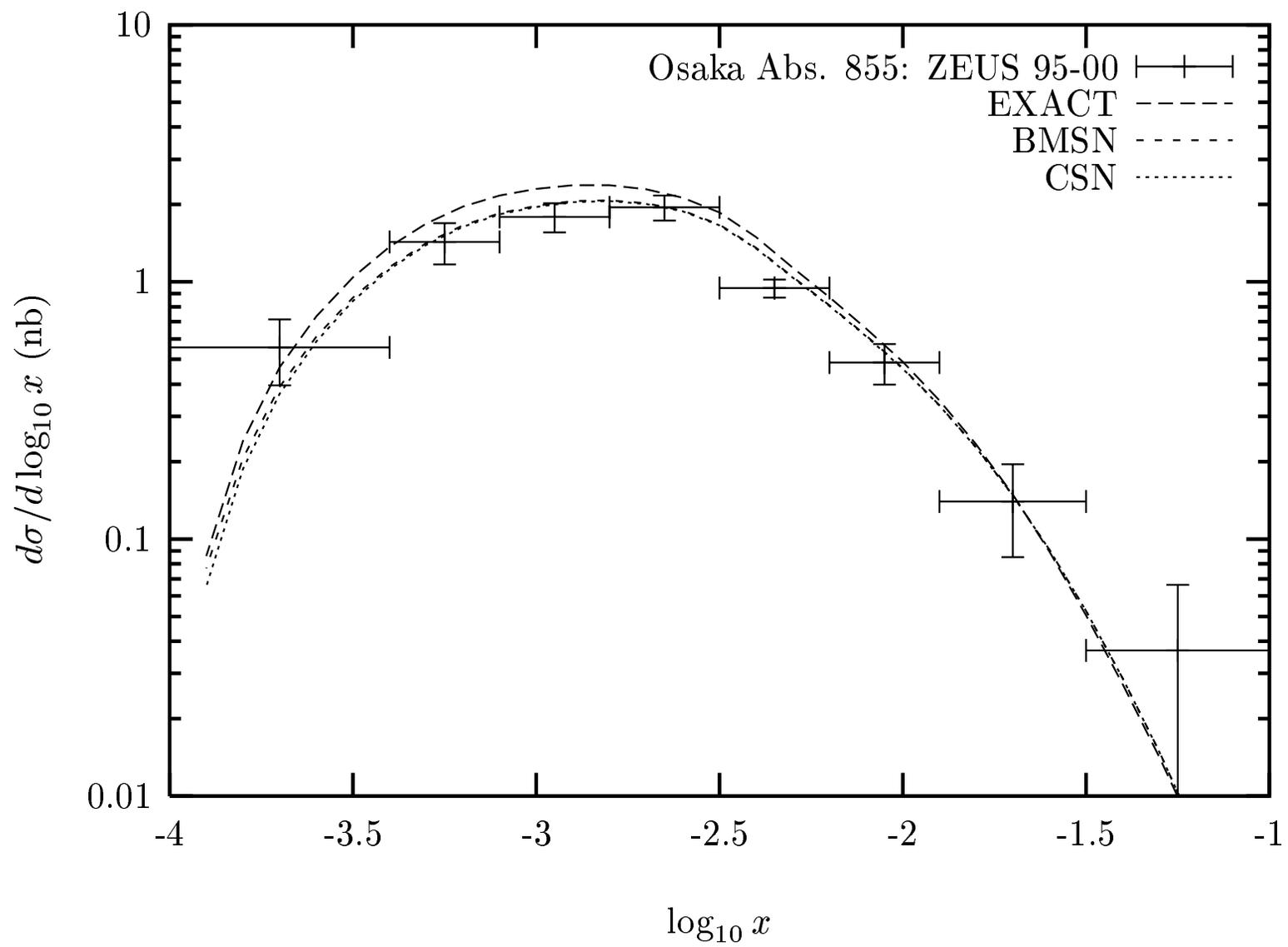
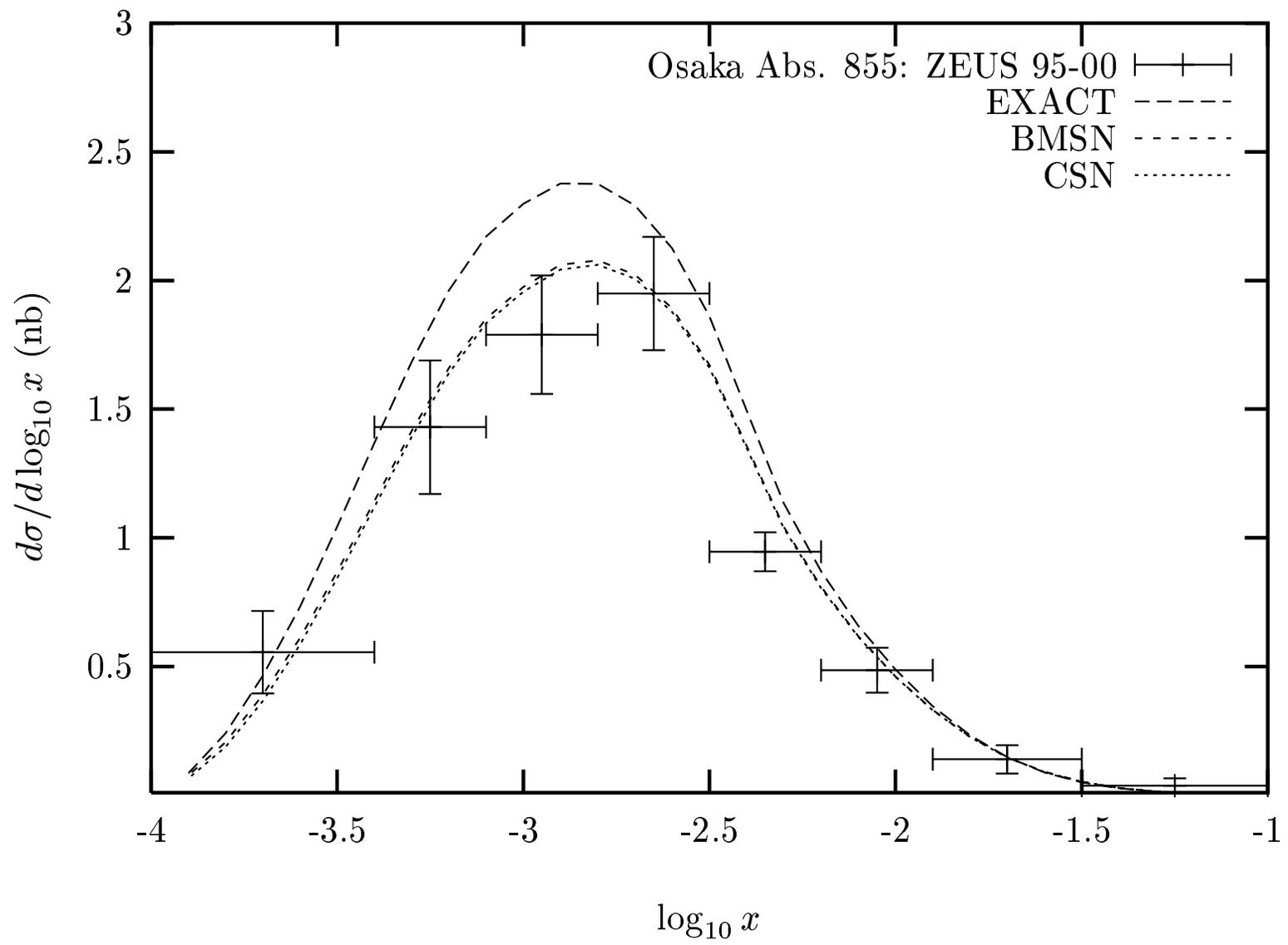
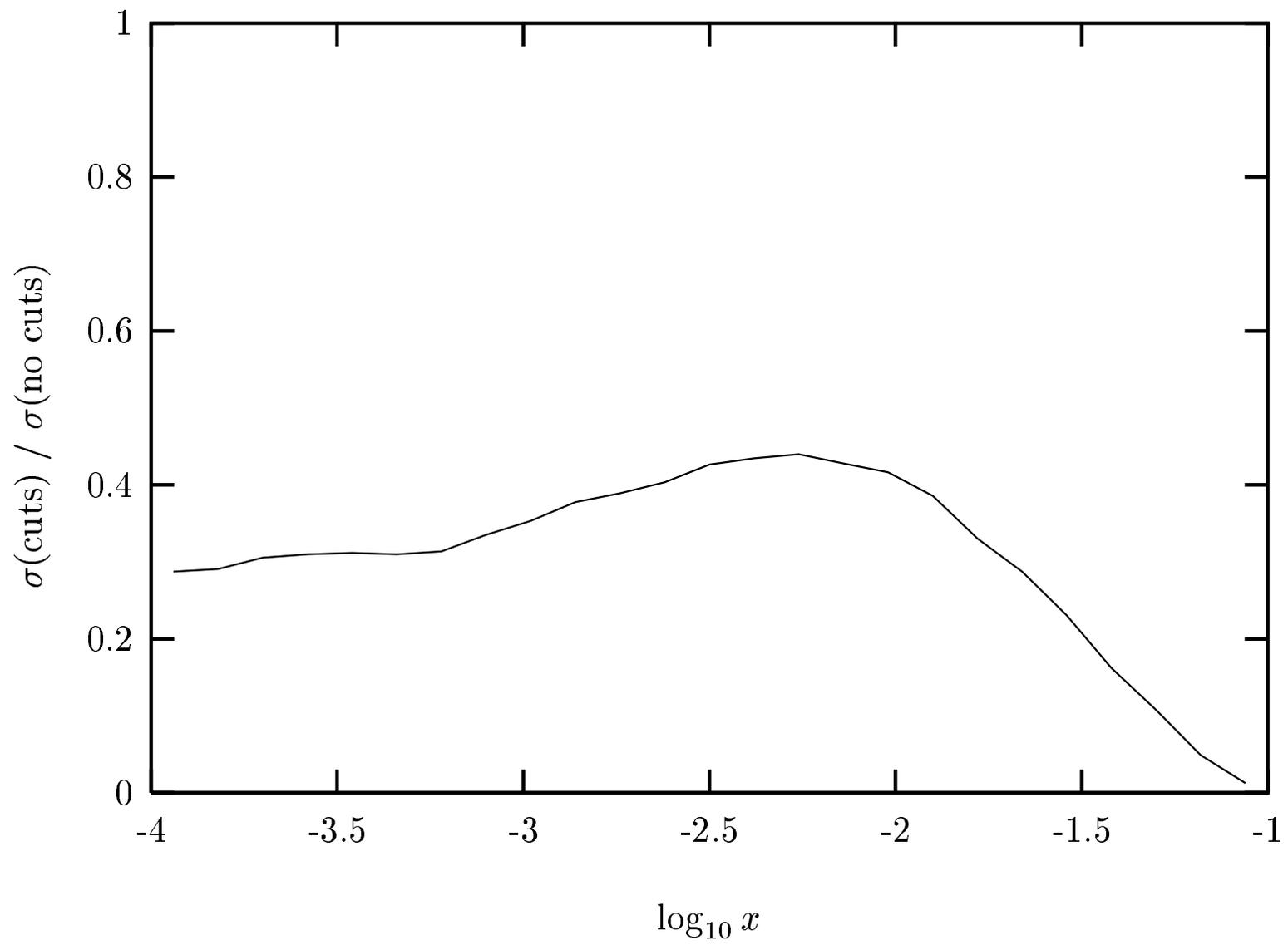
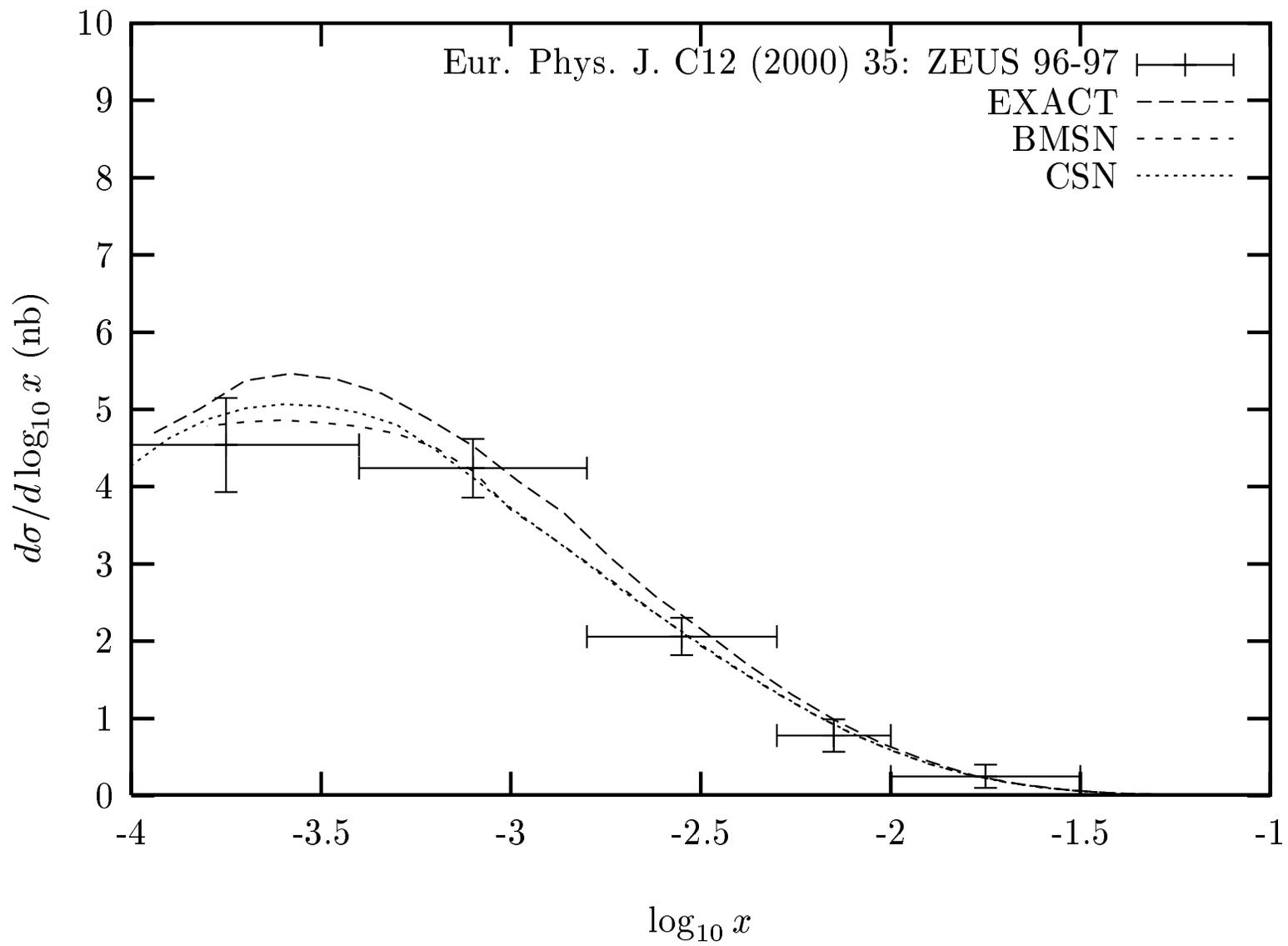
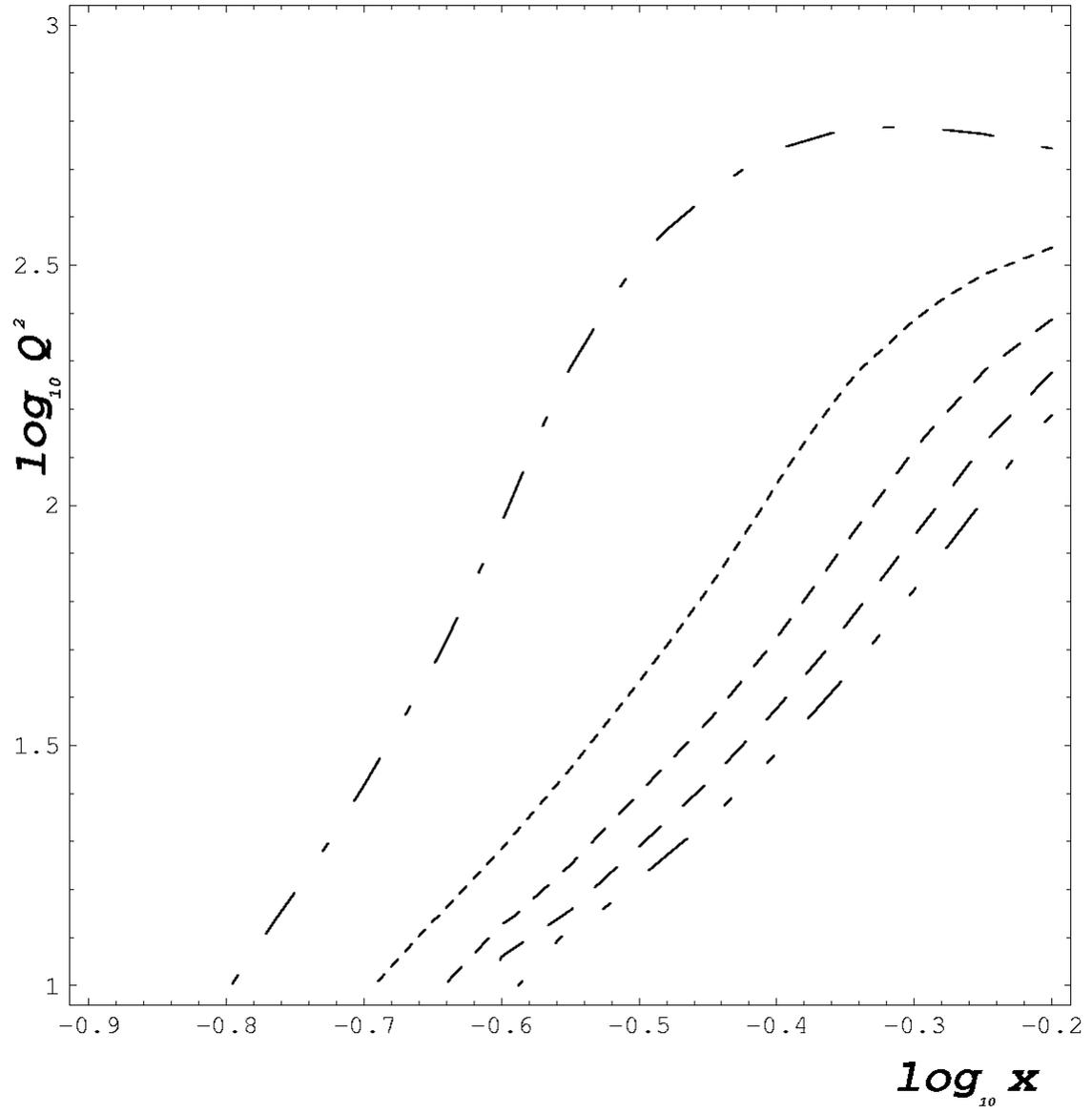
\end{document}